\documentclass[12pt]{iopart}

\usepackage{graphicx}
\usepackage{dcolumn}
\usepackage{bm}
\usepackage{color}

\begin{document}

\title[Effects of biaxial strain on the impurity-induced magnetism ...]{Effects of biaxial strain on the impurity-induced magnetism in 
       P-doped graphene and N-doped silicene: A first principles study}

\author{J. Hern{\'a}ndez-Tecorralco$^1$, L. Meza-Montes$^1$, M. E. Cifuentes-Quintal$^2$, and R. de Coss$^2$}

\address{$^1$ Instituto de F{\'\i}sica, Benem{\'e}rita Universidad Aut{\'o}noma de Puebla, 
             Apartado Postal J-48, 72570, Puebla, Puebla, M{\'e}xico}

\address{$^2$ Departamento de F{\'\i}sica Aplicada, Centro de Investigaci{\'o}n y de 
             Estudios Avanzados del IPN, Apartado Postal 73, Cordemex, 97310, 
             M{\'e}rida, Yucat{\'a}n, M{\'e}xico}
\ead{romeo.decoss@cinvestav.mx}

\date{\today}

\begin{abstract}

The effects of biaxial strain on the impurity-induced magnetism in P-doped
graphene (P-graphene) and N-doped silicene (N-silicene) are studied by means 
of spin-polarized density functional calculations, using the supercell approach. 
The calculations were performed for three different supercell sizes $4\times 4$, 
$5\times 5$, and $6\times 6$, in order to simulate three different dopant 
concentrations 3.1, 2.0 and 1.4\%, respectively. For both systems, the 
calculated magnetic moment is 1.0 $\mu_B$ per impurity atom for the three 
studied concentrations. From the analysis of the electronic structure and
the total energy as a function of the magnetization, we show that a Stoner-type 
model describing the electronic instability of the narrow impurity band accounts 
for the origin of $sp$-magnetism in P-graphene and N-silicene. Under biaxial 
strain the impurity band dispersion increases and the magnetic moment gradually
decreases, with the consequent collapse of the magnetization at moderate strain 
values. Thus, we found that biaxial strain induces a magnetic quantum phase 
transition in P-graphene and N-silicene.
\end{abstract}



\maketitle

\section{INTRODUCTION}

Magnetism is one of the most studied phenomena in physics and materials science. 
The magnetic behavior of the matter is usually attributed to the presence of 
\textit{d-} or \textit{f-} electrons. However, less common is the existence of 
magnetic materials with only \textit{s-} and \textit{p-} electrons. The arise of 
graphene \cite{novoselov} and related two-dimensional (2D) materials, such as silicene, 
have aroused great interest due to their outstanding properties and the expectation 
to exhibit $sp$-magnetism under certain conditions. Pristine graphene and silicene 
are non-magnetic, however, an alternative to induce magnetism is by introducing 
defects. For instance, it has been observed that vacancies \cite{yazyev} or 
the chemical functionalization using adsorbed \cite{yazyev,nair,gonzalez} 
or substitutional \cite{santos-1} impurities,  are effective ways to induce 
magnetism in 2D systems. Thus, the study of impurity-induced magnetism in 
low-dimensional materials is relevant in view of their potential applications 
in spintronics \cite{pesin,han} and spin-based quantum information systems
\cite{guido,guille}.

The electronic and magnetic properties of doped graphene and silicene with 
substitutional $sp$-impurities; Al, Si, P, and S, for graphene and  B, N, Al, 
and P for silicene, have been theoretically studied using first principles 
calculations based on the density functional theory \cite{dai-1,dai-2,wang,denis,sivek}.
Particularly, it has been reported that phosphorus and nitrogen atoms as single 
substitutional impurities in graphene (P-graphene) and silicene (N-silicene), 
respectively, present a net magnetic moment. Dai \textit{et al.} \cite{dai-1,dai-2} 
report a net magnetic moment of 1.05 $\mu_B$ in P-graphene for doping concentrations 
of 1.4 and 3.1 \%. The P atom introduces a local curvature in the graphene lattice 
and they report a metastable non-magnetic state when the P atom is at the plane. 
Wang \textit{et al.} \cite{wang} obtain similar results at a doping concentration 
of 2 \% with magnetic moment of 1.02 $\mu_B$. The authors attribute the origin 
of the magnetism to the symmetry breaking of $\pi$-electrons in graphene. 
Furthermore, they show that the spin density charge is distributed over the whole 
lattice. A systematic study of the concentration effect in graphene with 
$sp$-impurities as Al, Si, P, and S was performed by Denis \cite{denis}. 
These results show that for P-graphene the magnetic moment is independent of the 
concentration in a range of 0.8-3.1 $\%$. It is important to mention that within
the group of impurities Al, Si, P and S, only the P impurity induces magnetism 
in graphene.

For doped silicene, the effects of the chemical functionalization with B, N, Al 
and P impurities for a doping concentration of 3.1 $\%$ were analyzed by Sivek 
\textit{et al.} \cite{sivek}, using the density functional theory with the local 
density approximation for the exchange-correlation potential. Their results showed 
that the nitrogen substitutional 
impurity induces a magnetic moment of 0.9 $\mu_B$ and the system is vibrationally stable. 
The authors also discuss that the contribution of $s$- and $p$- states of N gives 
rise to metallic bands in silicene. Hence, in a similar way that for P-graphene, 
Sivek \textit{et al.} \cite{sivek} found that within the group of impurities 
B, N, Al, and P, only the N impurity induces magnetism in silicene. Thus, 
first-principles calculations predict that P-graphene and N-silicene belong to the 
group of $sp$-magnetic systems. However, the details of the mechanism that gives 
rise the magnetism in P-graphene and N-silicene still needs to be fully understood.

Additionally, having a technique that allows us to modulate the magnetic properties
of 2D materials is highly desirable. In this way, it has been showed that 
strain engineering is an effective method to modulate the electronic and magnetic 
properties in transition metal doped 2D systems \cite{li,zheng,santos-2}. Nevertheless, 
the study of strain effects on the impurity-induced $sp$-magnetism in P-graphene or 
N-silicene is still lacking. Thus, a systematic and comparative theoretical study of 
the magnetic properties in P-graphene and N-silicene is necessary in order to understand 
the effect of strain for different dopant concentration values.
  
Therefore, the aim of this work is to contribute to the understanding of the 
impurity-induced magnetism in P-graphene and N-silicene and how the magnetic moment of 
these systems could be modulated under a positive isotropic deformation (biaxial strain).
Here, we present results of first-principles calculations based on the Density Functional
Theory (DFT) for the structural, electronic and magnetic properties of 
P-graphene and N-silicene for three different doping concentrations (3.1, 2.0 
and 1.4\%) in a moderate range of deformations ($0-10\%$). Firstly, the first-principles
results for the electronic structure and the total-energy as a function of the
magnetization are analyzed using a Stoner-type model describing the electronic 
instability of a narrow impurity band, and we show that this model accounts 
for the origin of $sp$-magnetism in P-graphene and N-silicene. Secondly, the
results for the evolution of the magnetic moment as a function of the biaxial 
strain are presented. We find that biaxial strain gradually destabilizes the magnetic
state inducing a magnetic to paramagnetic phase transition in these systems. 
The paper is structured as follows: In Sec. II we describe the computational details 
of our calculations. In Sec. IIIA structural results are presented and the magnetic 
and electronic properties are discussed in Sec. IIIB. Finally, in Sec. IV we report 
our main conclusions. Particularly useful is Appendix A presenting a detailed 
description of the narrow impurity band model for ferromagnetism used throughout
the paper.

\section{Computational details}

The DFT calculations were performed within the framework of the plane-waves 
pseudopotential approach, as implemented in the QUANTUM-ESPRESSO code \cite{qe}. 
Core electrons were replaced by ultrasoft pseudopotentials taken from 
the \texttt{PSlibrary} 1.0.0 database \cite{pseudo} and the valence wave functions 
(charge density) were expanded in plane waves with a kinetic-energy cutoff of 55 
(340) Ry for graphene and 50 (320) Ry for silicene. The exchange-correlation 
functional was treated with the Perdew-Burke-Enzerhof \cite{pbe} parametrization 
of the generalized gradient approximation. Since our study systems involve only $s$ and $p$ electrons, we expect that this functional provides a good description of the magnetic behavior, compared with cases with highly localized $d$ or $f$ orbitals, where usually tends to give a poor description due to an effect of delocalization of the wavefunctions.
We simulated P (N) substitutional impurities by replacing one C (Si) atom from 
the graphene (silicene) pristine lattice. We considered 4$\times$4, 5$\times$5 and 
6$\times$6 supercells, with 32, 50, and 72 atoms which correspond to 3.1, 2.0 and 
1.4 \% of impurities concentrations $(c)$, respectively. For each concentration we 
calculate the corresponding ground-state lattice constant $a_0$ by a direct minimization 
of the electronic total energy. Biaxial tensile strain $\varepsilon$ was applied by 
increasing the lattice constant as $a=(1+\varepsilon)a_0$. During all the structural 
calculations the atomic positions were relaxed until the internal forces were less 
than 0.01 eV/\AA. In order to simulate an isolated layer, we left at least 15 \AA\ 
of vacuum space between periodic images. 
Special attention was paid to the sampling of the Brillouin zone. 
For structural calculations, we used a $9\times9$ k-grid \cite{kpoints} with a 
Methfessel-Paxton smearing of 0.015 Ry \cite{integration}. However, for electronic 
and magnetic properties we had to use a $18\times18$ $k-$grid with a smearing of 
0.005 and 0.002 Ry for graphene and silicene, respectively. This was needed in order 
to properly converge the magnetic moment up to 0.01 $\mu_B$.

\section{Results and discussion}

\subsection{Energetics and structural properties}

We begin our discussion by analyzing the energetic stability of the impurity on the host. For that, 
the binding energy for the unstrained ground state was calculated as 
\begin{equation}\label{eq01}
E_B=E_{system}-\left( E_{2D-vacancy}+E_{atom}\right)\, ,
\end{equation} 
where $E_{system}$ is the total energy of the full relaxed doped system, whereas 
$E_{2D-vacancy}$ is the total energy for a full relaxed vacancy and $E_{atom}$ 
is the total energy for the isolated impurity-atom, in this case P or N. For 
P-graphene we found for $E_B$ values of $-8.30$ eV, $-8.34$ eV and $-8.42$ eV, 
for the concentrations of 1.4, 2.0, and 3.1 \%, respectively, which are in good agreement with the value of $-8.31$ eV reported by Pa\v sti \textit{et al.} \cite{pasti}. In the case of 
N-silicene $E_B$ is $-7.24$ eV, $-7.25$ eV and $-7.22$ eV, for the concentrations
of 1.4, 2.0, and 3.1\%, respectively.  Thus, the calculated values for the binding energy show that the substitutional impurity of P(N) is energetically stable in graphene (silicene).

\begin{figure*}
\centering
 	\includegraphics[scale=0.9]{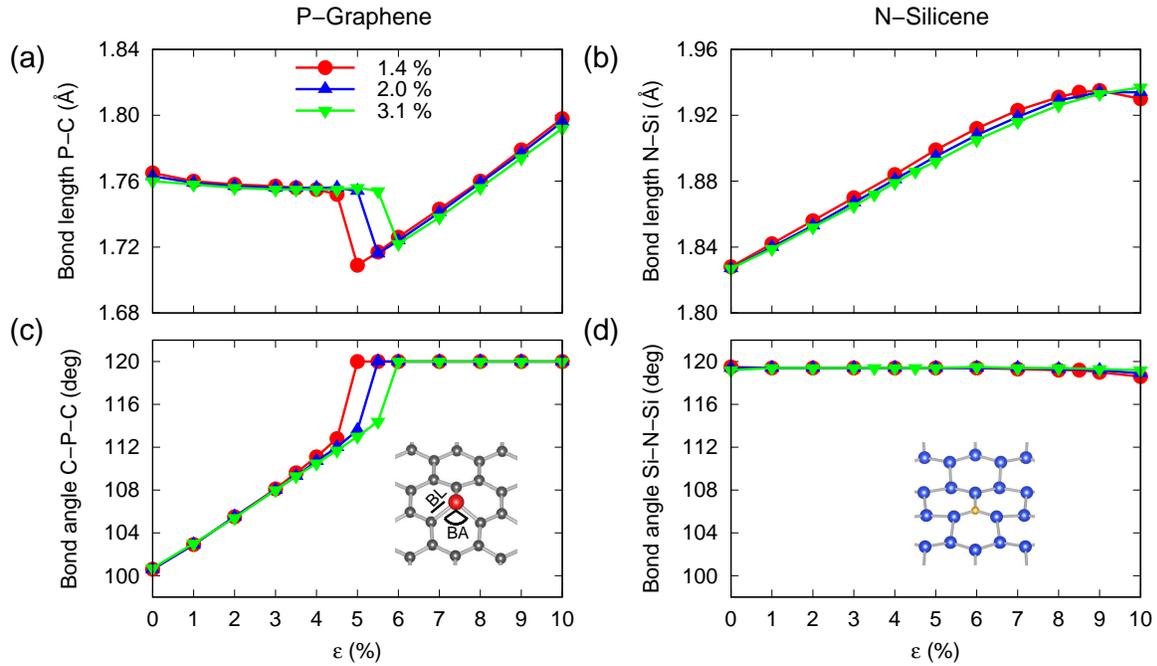}%
 	\caption{Bond length and bond angle for P-graphene(a, c) and N-silicene 
       (b, d) as a function of deformation $\varepsilon$, at the three studied
        concentrations (1.4, 2.0, and 3.1\%).}
 	\label{figure:fig1}
\end{figure*}

With respect to the structural properties, it is important to remember that pristine 
graphene is a flat crystal whereas pristine silicene has a buckled structure. Thus,
because of the different structural character and the contrasting size of the impurity
atom in each case, a different structural behavior for P-graphene and N-silicene is 
anticipated. The lattice structure for P-graphene in the ground (unstrained) state 
shows significant distortions owing to the presence of substitutional P, which has 
an atomic size larger than the carbon atom. The impurity causes a distorted 
tetrahedral-type structure with a bond angle of $\theta_{\mathrm{CPC}}=$100$^\circ$ 
and P-C bond length of 1.76 \AA\ which is larger than the C-C bond in pristine 
graphene (1.42 \AA). For N-silicene, because N atom is smaller in size than Si, as substitutional impurity it has a N-Si bond length smaller than Si-Si, 
given by 1.83 \AA\ and 2.27 \AA, respectively. The difference in distances causes 
a bond angle $\theta_{\mathrm{SiNSi}} = $119.4$^\circ$, which is close to the 
typical value of $sp^2$ hybridization whereas for pristine silicene the 
$\theta_{\mathrm{SiSiSi}} = $116.2$^\circ$ is usually attributed to $sp^2$-$sp^3$ 
hybridization. The main structural difference is localized around the impurity, 
for P-graphene the P atom is out of the plane respect to the flat graphitic 
lattice whereas for N-silicene the N atom is in the same plane with their three 
first nearest neighbors of a deformed blucked structure. That is the reason for 
the different evolution of the structural parameters as a function of the biaxial 
strain for each system as it can be seen in Fig.~\ref{figure:fig1}. 

For both systems, the evolution of the bond length (BL) and the bond angle (BA) 
is almost independent of the concentration, but under biaxial deformations 
($\varepsilon$), an appreciable difference occurs between P-graphene and N-silicene 
as we discuss below. In the case of P-graphene we can observe two regimes 
(see Fig.~\ref{figure:fig1}a), for 0 to 5\% of strain the BL is almost constant 
while P is completely out of the plane, followed by a reduction and then a linear 
increment from 5 to 10\%. From 0 to 5\% the BA increases with deformation and from 
5 to 10\% the angle becomes constant $\theta = 120^\circ$ (see Fig.~\ref{figure:fig1}c), 
indicating that the system becomes flat as pristine graphene for deformations larger 
than 5\%. For N-silicene, in Fig.~\ref{figure:fig1}b we can see that the BL 
has a monotonic increase with strain, although around of 9\% it seems to reach 
a maximum. With respect to the BA in N-silicene under strain, interestingly, 
we can see in Fig.~\ref{figure:fig1}d) that biaxial strain does not affect the 
$\theta_{\mathrm{SiNSi}}$, remaining almost constant in the whole range of
deformations 0-10\%. This means that for the N atom is more favorable to be 
almost aligned to their first nearest neighbors. This behavior is an effect 
of the strong hybridization between the N and Si orbitals, avoiding a buckled 
configuration around N.

\subsection{Electronic and magnetic properties}

In carbon-based magnetism, it has been proposed that the origin of $sp$-magnetism 
can be explained by the Stoner theory adapted to the case of a narrow impurity
band \cite{edwards}, where a paramagnetic system becomes unstable with respect to 
the ferromagnetic case when a high density of states exists at the Fermi level. 
The Stoner criterion for the existence of ferromagnetism is $I N(E_F) > 1$, where 
$I$ is the Stoner parameter and $N(E_F)$ is the density of states (DOS) of the 
paramagnetic case at Fermi level. It has been suggested that this criterion is
useful to describe the emergence of magnetism in systems with $sp$-electrons,
particularly in graphene nanostructures with vacancies or with adsorbed hydrogen 
atoms \cite{yazyev,gonzalez,yazyev-rpp,kwlee}. The cases of substituted graphene 
and even silicene are not the exception. However, for these substitutional cases, 
the role of the specific impurity is important, otherwise, other $sp$-impurities 
could induce magnetism in graphene and silicene. An alternative to the Stoner 
model for the ferromagnetism occurring in a narrow impurity band can be found in 
Appendix A, which is based on the proposal of Edwards and Katsnelson \cite{edwards} 
and the work of Gruber {\it et al.}\cite{mohn-model}

\begin{figure*}
\centering
      \includegraphics[scale=1.0]{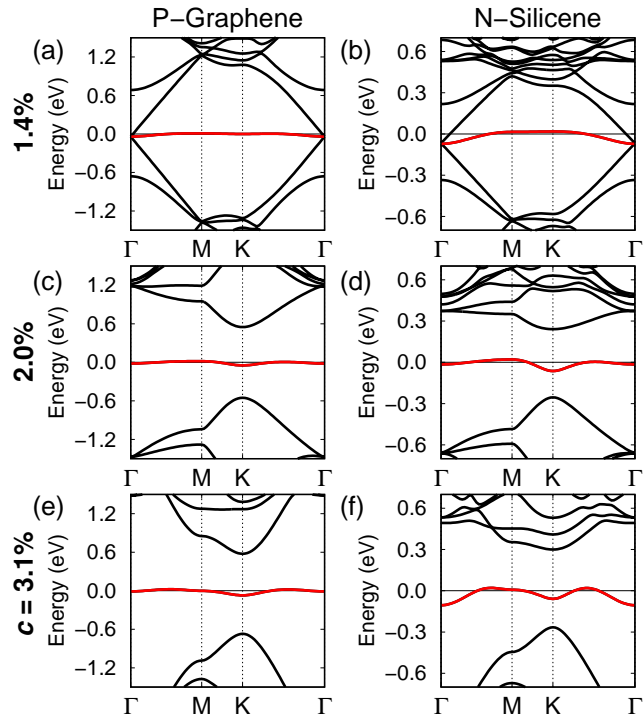}
      \caption{Electronic band structure for P-graphene and N-silicene in the
               paramagnetic state for the three studied concentrations. The origin
               of the energy scale has been set at the Fermi level $(E_F)$. The 
               narrow impurity band is emphasized in red.}
     \label{figure:fig2}
\end{figure*}

In Fig.~\ref{figure:fig2} the paramagnetic electronic band structure of 
P-graphene (left) and the N-silicene (right) are shown at the unstrained state
for the three different values of concentration ($c$ = 1.4, 2.0 and 3.1\%). As it 
is well known, the band structure of graphene and silicene shows a linear 
dispersion around the Fermi level in the K point, the so-called Dirac cones. 
From Fig.~\ref{figure:fig2}, it is observed that for concentrations of 
2.0 and 3.1\% this linearity dissapears with doping, and in fact we find a
band gap opening and a narrow impurity band at the Fermi level (band in
red color). For the concentration of 1.4\% the band structure shows differences with respect to the other two concentrations due to band folding effect by the use of the supercells \cite{yzhou}.  For this concentration, the K point of the unit cell is folded to the $\Gamma$ point of the supercell as Figs. 2a and 2b show. Besides, the band gap opening does not occur, but the impurity band is present. It is interesting to note that the
dispersion of the impurity band is larger in N-silicene with respect to
P-graphene. Nevertheless, in both systems the dispersion of the impurity
band increases with the concentration. A further characterization of the 
impurity band was done and we find that for both systems the Fermi level 
is at half-filling and that the maximum occupancy is 1.0. Additionally, 
the value of the impurity bandwidth ($W_{imp}$) for the concentration of 2.0\% 
in P-graphene and N-silicene is 65 and 84 meV, respectively. This very 
narrow impurity-band is able to produce a sharp peak in the Density of States 
(DOS) at Fermi level. Hence, an electronic instability in the paramagnetic state 
is expected from the high value of $N(E_F)$, in particular a band-splitting 
induced by spin polarization, generating a net magnetic moment in the system. 

\begin{figure*}
\centering
      \includegraphics[scale=1.0]{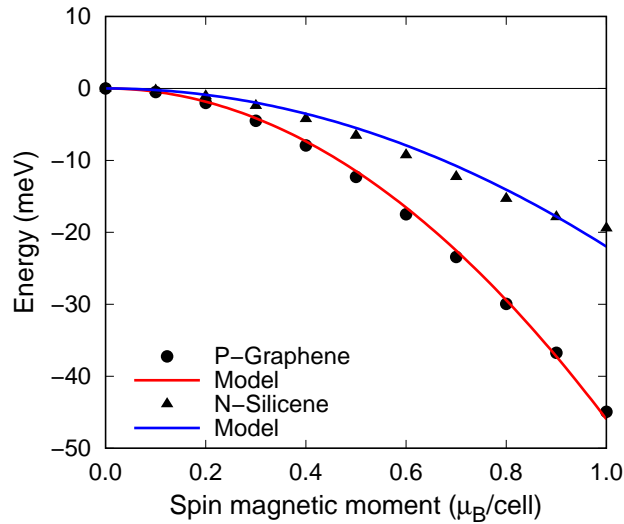}
      \caption{Energy as a function of the magnetic moment for P-graphene
               and N-silicene at $c=2\%$, calculated using the fixed-spin 
               moment method (symbols). The solid line corresponds to the 
               fit of $E(M)$ given by eq. (\ref{eq02}) to the calculated 
               values (symbols).}
     \label{figure:fig3}
\end{figure*}

In order to have more physical insight on this instability, we have performed
first-principles calculations for the total-energy as a function of the magnetic 
moment in the supercell using the Fixed Spin Moment (FSM) method \cite{mohn-fsm}. 
In Fig.~\ref{figure:fig3}, we show the calculated values (symbols) of the
total-energy as a function of the spin magnetic moment, $E(M)$, for P-graphene 
and N-silicene for $c=2.0\%$, using as a reference the total-energy of the 
paramagnetic case. In both cases the ferromagnetic state has lower energy than 
the paramagnetic state. The behavior of $E(M)$ indicates that the most stable 
state is the full polarized state with a spin magnetic moment of 1.0 $\mu_B$/cell, 
corresponding to strong ferromagnetism \cite{friedel,mohn-book}.

From the model for ferromagnetism in a narrow impurity band described in Appendix A,
we have that in the case of half-filling, maximum occupancy $N_0=1$, and assuming
the rigid band splitting, the total energy as a function of the magnetic moment is
thus given by
\begin{equation}\label{eq02}
E(M) = E_0 + \frac{M^2}{4}\left(W_{imp} - U \right),
\end{equation}
where $E_0$ corresponds to the reference energy, for instance the total energy of
the paramagnetic state, and $U$ to the Coulomb type interaction. Thus, the condition
for spontaneous magnetization is $U/W_{imp}>1$ \cite{friedel,mohn-model,mohn-book}.  
The values for $W_{imp}$ where obtained previously from the paramagnetic band structure, 
65 meV for P-graphene and 84 meV for N-silicene at $c=2.0\%$. Hence, to obtain the values 
for $U$, we have fitted (\ref{eq02}) to the calculated values of $E(M)$, solid line in 
Fig.~\ref{figure:fig3}, resulting in $U=250$ meV for P-graphene and $U=172$ meV for 
N-silicene. Therefore, $U/W_{imp}$ is 3.8 and 2.0 for P-graphene and N-silicene, 
respectively, fulfilling the Stoner-type criterion $U/W_{imp}>1$. From this quantitative 
analysis of the total-energy as a function of the spin magnetic moment using a Stoner-type 
model, it is clear that the value of $U$ in these systems is small. Consequently, the 
spontaneous polarization in P-graphene and N-silicene is driven by the very small value 
of $W_{imp}$. 

The electronic band structure and DOS for the spin-polarized state for P-graphene and 
N-silicene are presented in Figures 4 and 5, respectively. The spin-up states are 
indicated in red color and the spin-down states in blue color. For reference, also
we have included the paramagnetic bands in grey color.  The largest spin-splitting
is for the impurity band, but the host bands also show an important spin-splitting,
indicating that the spin-polarization is not only localized at the impurity atom.
It is also interesting to note that the impurity bandwidth for spin-down 
is larger than for the spin-up, indicating that the spin-polarization of the impurity
band is not a rigid-band splitting. The impurity band splitting is a result of
the exchange interaction between the electrons in the narrow impurity band. Thus,
in order to characterize the exchange interaction, we have calculated the spin-splitting
of the impurity band $(\Delta_s)$ by taking a weighted average in the first Brillouin 
zone. The values for $\Delta_s$ in P-graphene and N-silicene are 267 meV and 137 meV,
respectively. 

\begin{figure*}
\centering
      \includegraphics[scale=0.9]{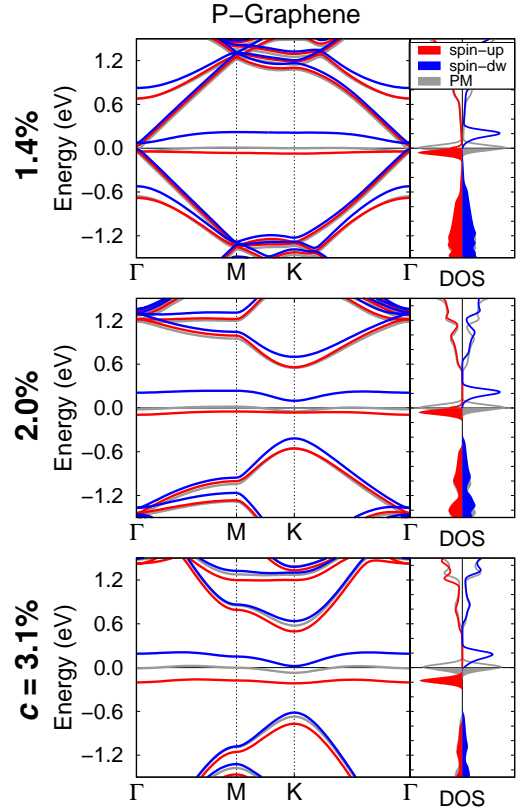}
      \caption{Spin-polarized electronic band structure for P-graphene at the 
               three studied concentrations. The red and blue lines correspond
               to the spin-up and spin-down, respectively. For reference, the
               paramagnetic case is included in grey color.}
     \label{figure:fig4}
\end{figure*}

\begin{figure*}
\centering
      \includegraphics[scale=0.9]{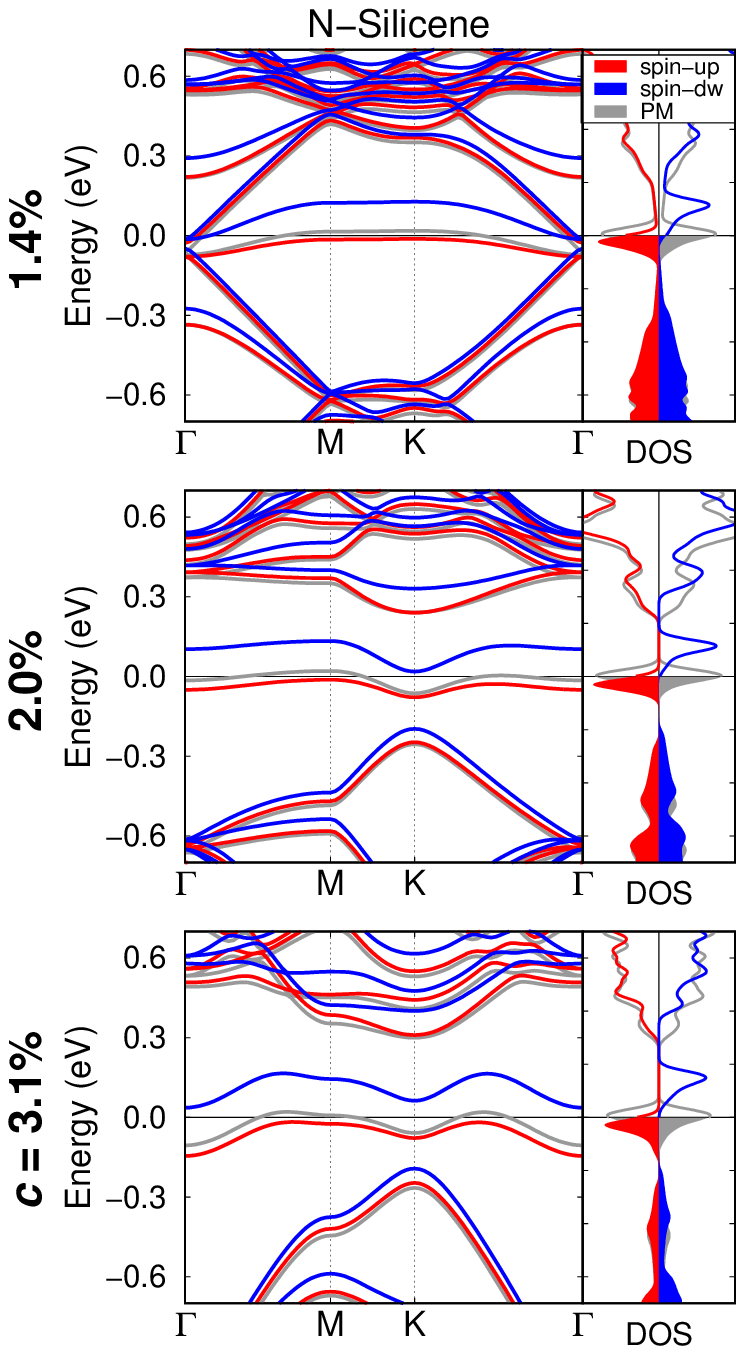}
      \caption{Spin-polarized electronic band structure for N-silicene at the 
               three studied concentrations. The red and blue lines correspond
               to the spin-up and spin-down, respectively. For reference, the
               paramagnetic case is included in grey color.}
     \label{figure:fig5}
\end{figure*}

The analysis of the density of states revealed that the high DOS at Fermi level 
in Figures 4 and 5, which gives rise to magnetism comes from the hybridization of 
the orbitals at the impurity atom and the orbitals at the atoms which belong 
to the sublattice adjacent to the impurity. For P-graphene the localized state at 
the Fermi level have contributions of $s$ and $p_z$ orbitals from P, and from $p_z$ 
orbital of C atoms. In the case of N-silicene, the main contributions to the peak
at the DOS that causes the instability come from the $s$ and $p_z$ orbitals from N, 
$s$ and $p_z$ from Si, and a small contribution $p_x$ and $p_y$, which makes sense 
due to the buckled structure of silicene. 

\begin{figure*}
\centering
	\includegraphics[scale=1.0]{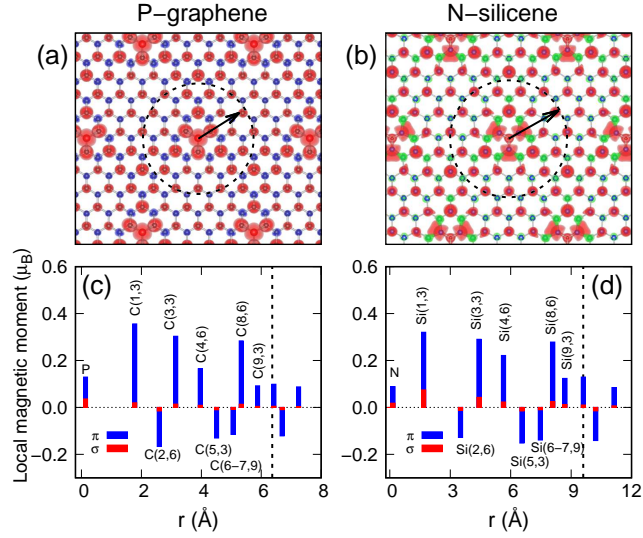}
	\caption{Spin density charge (top) and the $\pi$ and $\sigma$ contributions 
               to the local magnetic moments as a function of radial distance from
               the impurity atom (bottom) for P-graphene and N-silicene at $c=2\%$.
               The two indexes at each C or Si atom position represents the order 
               of nearest neighbor with respect to the impurity and the number of 
               nearest neighbors.}
	\label{figure:fig6}
\end{figure*}

As a first approximation to the analysis of the local magnetic moments distribution, 
we analyzed the L\"owdin charges for each atom. On the top of Fig.~\ref{figure:fig6} 
we have shown iso-surfaces of the spin electronic density charge 
($\rho^\uparrow - \rho^\downarrow$) whereas the local magnetic 
moment distributions are plotted at the bottom. The red color corresponds to 
majority spin density and blue (green) color to minority spin density for
P-graphene (N-silicene). For the bottom of Fig.~\ref{figure:fig6}, the color 
represents the contribution to the local magnetic moment from $\sigma$ ($s$+$p_x$+$p_y$) 
orbitals in red and $\pi$ ($p_z$) orbitals in blue. The spin density charge plots 
are presented in an extended supercell which involves a supercell with its first 
six neighbors. We can observe a spin distribution spread on all of the lattice, 
where the impurity and their first atomic three neighbors have a majority spin 
density. Interestingly, each kind of spin density is majority on each sublattice. 
The dashed black line is used to define a radial distance with respect to the impurity. 
Inside this circumference, we plot the local magnetic moments as a function of the 
radial distance and for each distance, we have performed a sum according to the 
number of the atoms at that distance. It is clear that the local magnetic moment 
distribution is not homogeneous on the lattice, but the pattern is the 
same in both P-graphene and N-silicene. Interestingly, only a small fraction of 
the total magnetic moment is located at the impurity atom. 

Thus, from the present analysis we can now answer the question; why P-graphene and 
N-silicene are magnetic? According to our observations, there are two fundamental 
reasons: {\it i}) the impurity have an electronic configuration with one more 
electron than the host material, and {\it ii}) the impurity induces a very
narrow band at Fermi level.


After the analysis and characterization of the electronic structure and the 
spin-magnetic moment for unstrained P-graphene and N-silicene, in Fig. 7,
we show the evolution of the electronic structure with strain for P-graphene
and N-silicene at $c=2.0\%$ in the paramagnetic state. The band structure along 
the high symmetry paths of the first Brillouin zone in P-graphene shows a narrow 
band around the Fermi level, except in the K point where a parabolic dispersion 
is observed. As the strain increases, this effect is more pronounced. The electronic 
behavior in N-silicene under isotropic deformation shows a different evolution. 
The nearly flat impurity band around the Fermi level along the high symmetry paths 
with a small parabolic dispersion around the K point observed for $\varepsilon=0\%$,
is strongly distorted under biaxial strain. As the deformation is increased the 
flat character disappears and shows parabolic dispersion around the K and $\Gamma$ 
points. Thus, we find that in both systems the dispersion of the impurity
band increases with the biaxial strain.

\begin{figure*}
\centering
	\includegraphics[scale=0.9]{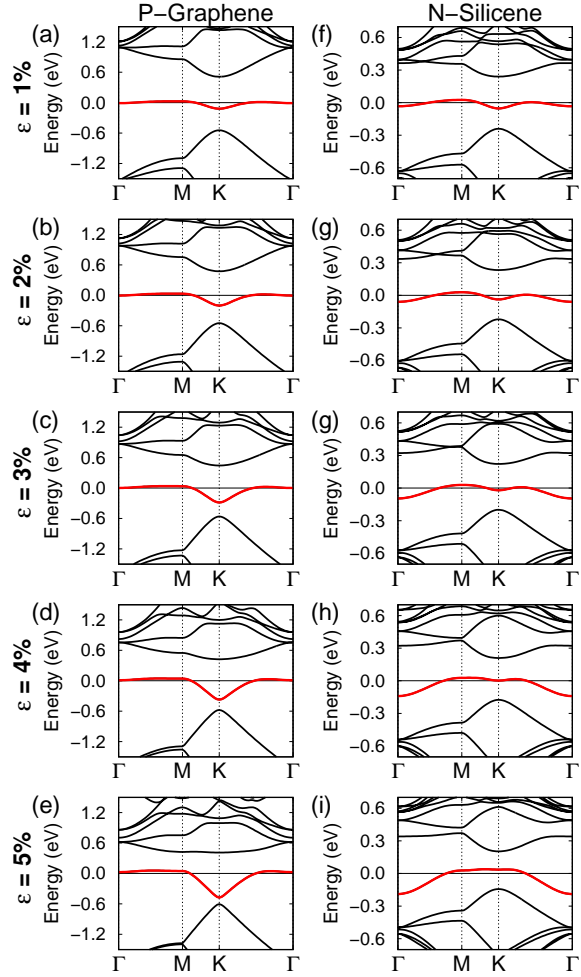}
	\caption{Electronic band structure for P-graphene and N-silicene at $c=2.0\%$
               in the paramagnetic state as a function of the strain. The origin
               of the energy scale has been set at the Fermi level $(E_F)$. The 
               impurity band is emphasized in red color.}
	\label{figure:fig7}
\end{figure*}

In Fig. 8 we show the behavior of $W_{imp}$ as a function of the biaxial strain
for P-graphene and N-silicene for the three studied concentrations, as obtained from
the paramagnetic band structure. For P-graphene, we can see that $W_{imp}$ follows 
a linear behavior with the applied strain in the range 0-5\% of deformations
for the three concentrations, with a large step at $\varepsilon\sim 5\%$. For
N-silicene, the behavior of $W_{imp}$ with the biaxial strain is strongly dependent 
on the concentration. For instance, for $c=3.1\%$ the impurity bandwidth increases
monotonically with the deformation, while for $c=1.4\%$ the $W_{imp}$ shows a 
minimum around 4\% of deformation. According to the narrow impurity band
model discussed above, these results anticipate the loss of magnetism in both 
systems, but the behavior of the magnetic moment with the biaxial strain and the 
concentration dependence for each system will be different.

\begin{figure*}
\centering
	\includegraphics[scale=1.0]{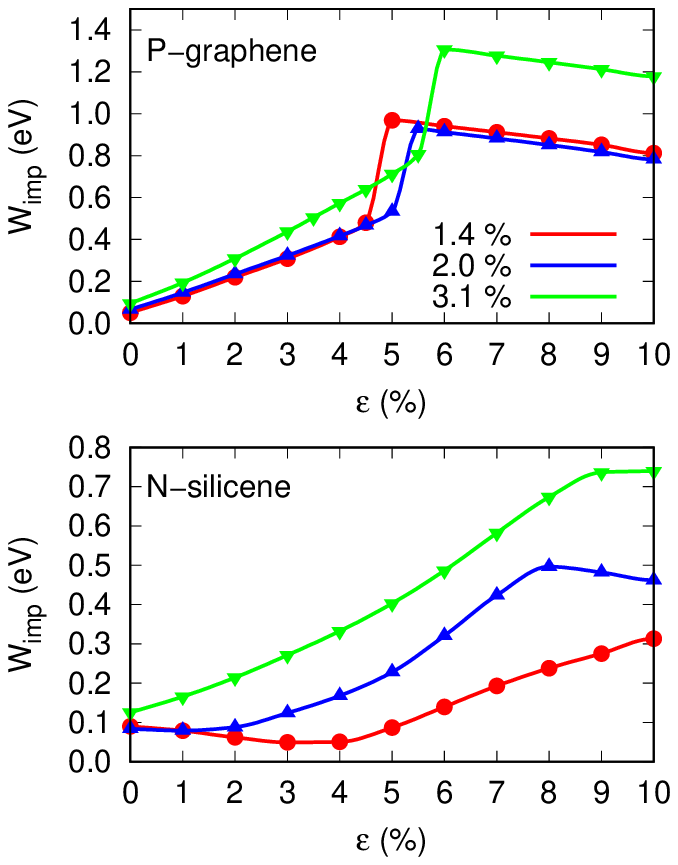}
	\caption{Evolution of the impurity bandwidth ($W_{\rm imp}$) as a
               function of the strain ($\varepsilon$) for P-graphene and 
               N-silicene at the three studied concentrations (1.4, 2.0, 
               and 3.1\%) in the paramagnetic state.}
	\label{figure:fig8}
\end{figure*}

In Fig. 9, the spin-polarized electronic band structure and DOS for P-graphene 
and N-silicene at $c=2.0\%$ under biaxial strain are presented. In both systems
we find that the impurity band dispersion increases with the strain for both
spin channels, inducing overlapping of the spin-up and spin-down bands, as can
be seen in the DOS. At the same time, the spin-splitting of the impurity band 
decreases. For this concentration, we can see that in both systems the spin-splitting 
vanishes at $\varepsilon=6\%$, indicating that the magnetism has been lost.

\begin{figure*}
\centering
	\includegraphics[scale=0.8]{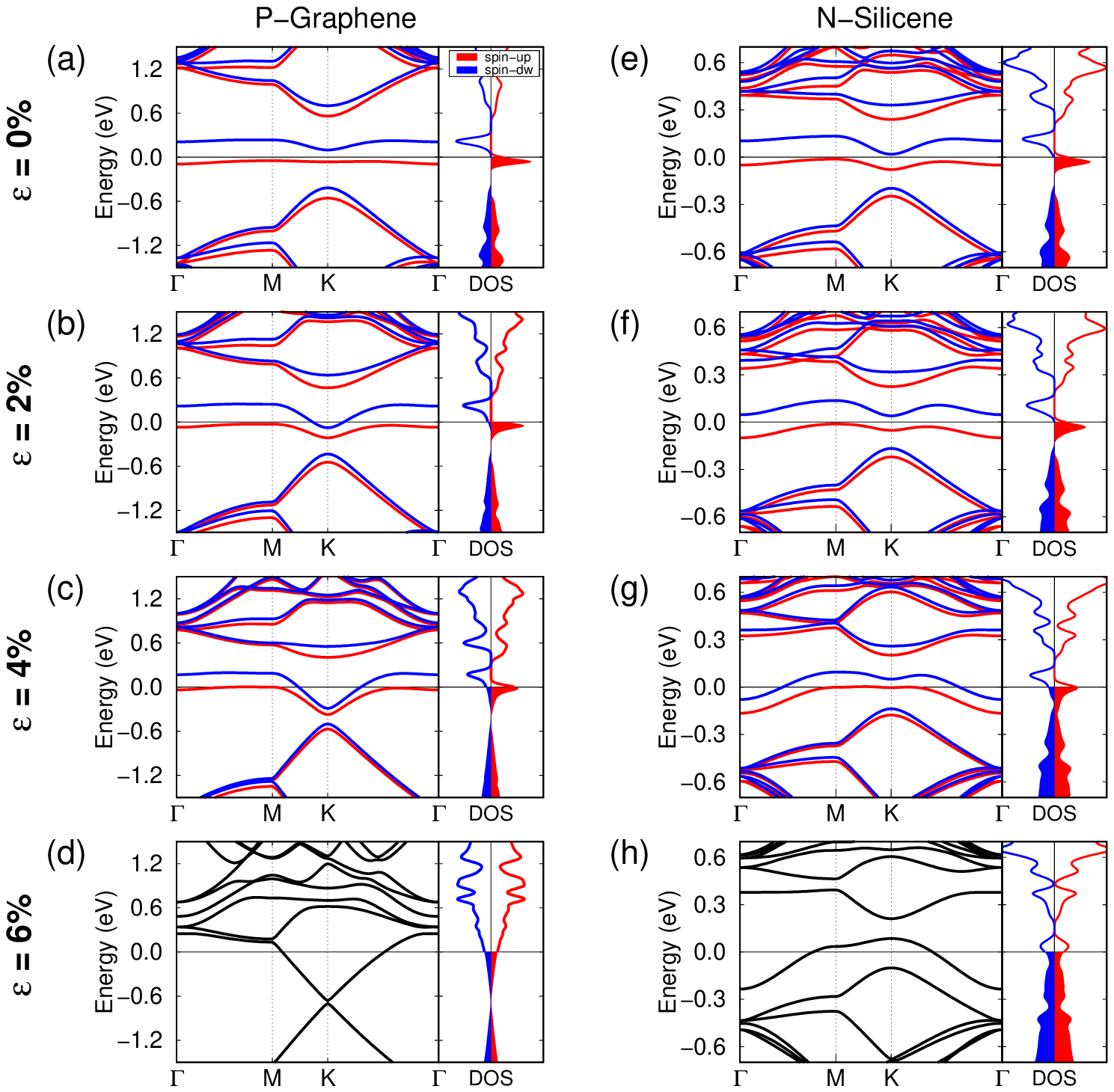}
	\caption{Spin-polarized electronic band structure for P-graphene and 
               N-silicene at $c=2.0\%$ as a function of the strain. The origin 
               of the energy scale has been set at the Fermi level $(E_F)$.}
	\label{figure:fig9}
\end{figure*}

Interestingly, for P-graphene at $\varepsilon=6\%$ we recover a linear dispersion 
around K point as in graphene pristine (Fig. 9d), however, the Fermi level is located 
above the Dirac point corresponding to electron doped graphene. For this deformation, 
as mentioned before, the system recovers its flatness, showing the close relation 
between structural and electronic properties. The electronic structure of N-silicene 
under isotropic deformation (Fig. 2e-h) shows a different evolution with respect to 
P-graphene. In this case, the impurity band at $\varepsilon=0\%$ shows a parabolic 
dispersion around the K point and a nearly flat character along the high symmetry paths. 
However, with deformation the evolution of the impurity band leads to a parabolic nature 
around K and $\Gamma$ points, with hole character at K and electron character at 
$\Gamma$. We notice that for N-silicene under strain, in contrast to P-graphene,
we do not recover the Dirac cones (see Fig. 9h). Thus, the electronic character
of N-silicene under biaxial strain in the non-magnetic state will be of a normal
metal.

\begin{figure}
\centering
	\includegraphics[scale=1.0]{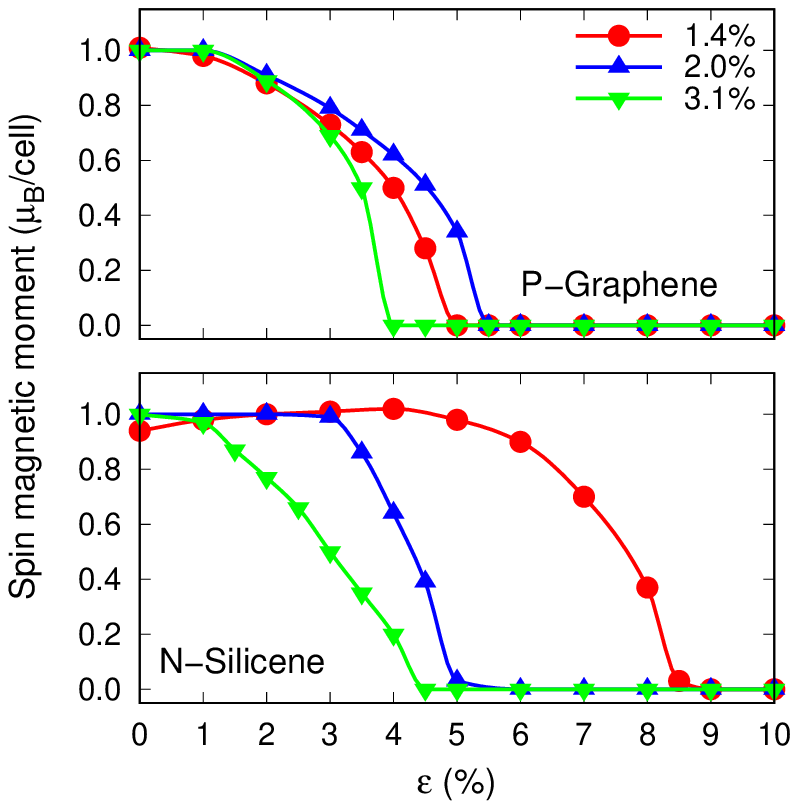}
	\caption{Spin magnetic moment as a function of biaxial strain for P-graphene 
               and N-silicene at the three studied concentrations (1.4, 2.0, 
               and 3.1\%).}
	\label{figure:fig10}
\end{figure}

The evolution of the magnetic moment as a function of the strain for each 
concentration is shown in Fig.~\ref{figure:fig10}. For the unstrained case 
both systems are magnetic with a net magnetic moment of 1.0 $\mathrm{\mu_B/cell}$ 
regardless of concentration. Under strain, the magnetic moment changes from 
1.0 to 0 $\mathrm{\mu_B/cell}$, indicating that a magnetic transition appears 
when a biaxial strain is applied. Although P-graphene presents almost similar 
transition with the doped concentration, N-silicene has a strong dependence
on the doping concentration. Nevertheless, a common feature is that in both
systems we have a range of deformations starting from $\varepsilon=0\%$
where the magnetic moment remain constant ($M=1.0$), a second range where the 
magnetic moment begins to decrease with strain ($0 < M < 1.0$) and a third range 
where the system is non-magnetic ($M=0$), after reaching the critical deformation
where $M\to 0$.

In order to understand the behavior of the magnetic moment in P-graphene 
and N-silicene with the applied strain, we return to the model of ferromagnetism
in a narrow impurity band, described in detail in the Appendix A. To simplify the analysis we will assume that the value of $U$ does not change with strain. This consideration comes from a further analysis of the spin-splitting for the impurity band and the magnetic moment as a function of the strain, which shows a practically constant behaviour of $U$ at different deformations.
In this case both systems are at half-filling and the maximum occupancy is $N_0=1$,
therefore the full polarized state remains as long as $W_{imp}<U$. This condition is 
fulfilled in the first range where the spin magnetic moment is 1.0 $\mu_B$/cell 
with the system in the strong ferromagnetic state.  In Fig. 8, we can see that for 
P-graphene the impurity bandwidth $W_{imp}$ increases linearly with strain
independently of the doping concentration, with practically the same slop up to 
$\varepsilon\sim 5\%$. This is the transition region where the system is in
the weak ferromagnetic state, characterized by an unsaturated spin-up band
and a magnetic moment $M < 1.0 \mu_B$/cell. In the case of N-silicene, the
picture is the same, however the details of the transition are strongly
dependent on the doping concentration as a result of the different behavior
of $W_{imp}$ with strain (see Fig.8). For the analysis of N-silicene, we will 
start with the highest concentration $c=3.1\%$, where the $W_{imp}$ increases
almost linearly with the biaxial strain. Thus, in this particular case the 
transition of the magnetic moment is similar to P-graphene. For N-silicene at 
$c=2.0\%$ the value of $W_{imp}$ remain almost constant up to $\varepsilon\sim 3\%$,
extending the first region corresponding to the strong ferromagnetic state,
but with a narrower transition region. Finally, for N-silicene at $c=1.4\%$ we
can see in Fig. 8 that $W_{imp}$ decreases in the range of 0-4\% for the strain,
reaching a minimum value at $\varepsilon=4\%$, and then increase linearly with 
a very small slope. In this way, the strong ferromagnetic state in N-silicene
at $c=1.4\%$ extends up to $\varepsilon=4\%$ and the wide transition region up 
to $\varepsilon=8.5\%$.

To conclude, despite the differences in the behavior of the magnetic moment 
with the biaxial strain for P-graphene and N-silicene, in particular the strong 
dependence in the case of N-silicene with the doping concentration, a common 
feature was recognized and characterized, emerging the picture depicted in
Fig. 11. So, we have that for low strain values the system remains in the 
strong ferromagnetic (SF) state, followed by a transition region corresponding 
to a weak ferromagnetic (WF) state with a impurity band partially polarized, 
and finally the non-magnetic region after reaching the critical strain where 
the ground state corresponds to a paramagnetic (PM) state. This picture, 
was confirmed performing fixed spin moment calculations of the total energy 
as a function of the spin magnetic moment, for three values of biaxial 
deformations 0, 4.0 and 7.0\% corresponding to SF, WF, and PM states, 
respectively. The results are presented in the inset of Fig. 11. 
The calculated plots for $E(M)$ are in close agreement with the 
general model of itinerant electrons in a narrow band \cite{mohn-book} .

\begin{figure*}
\centering
      \includegraphics[scale=1.0]{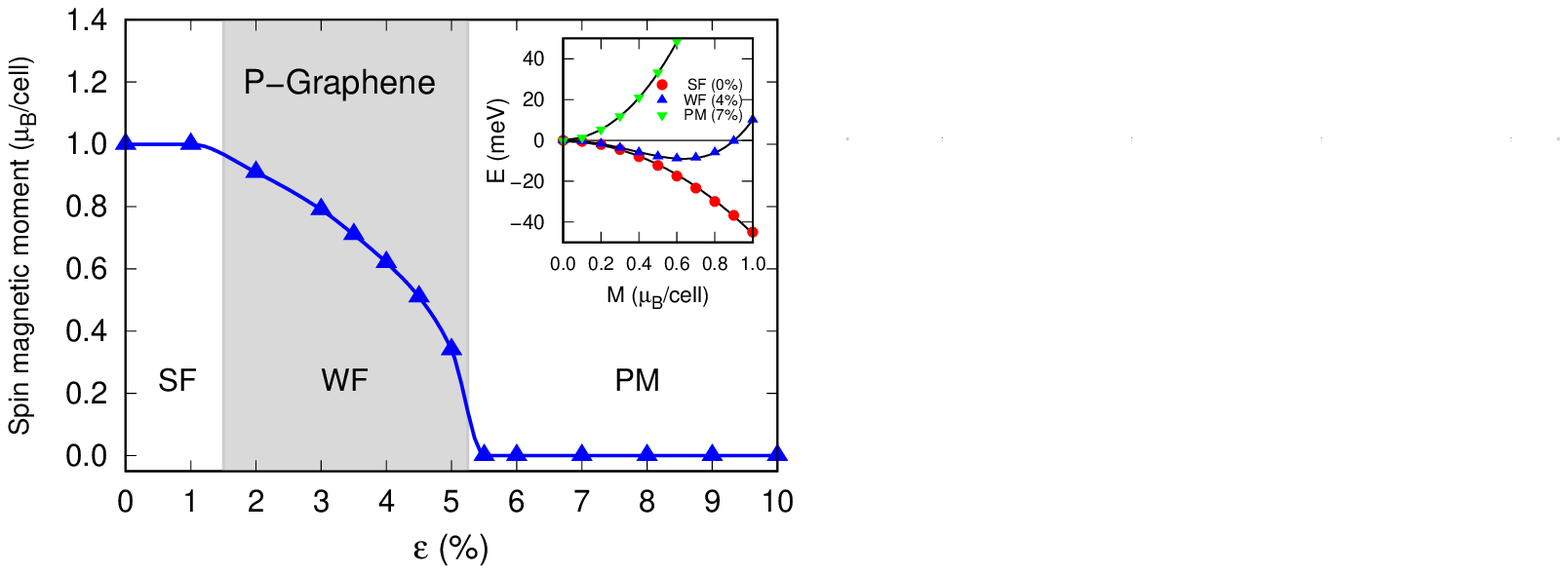}
      \caption{Evolution of the spin-magnetic moment as a function of the
               strain for P-graphene at $c=2\%$ indicating the regions for
               strong-ferromagnetism (SF), weak-ferromagnetism (WF) and 
               paramagnetic (PM) behaviors. In the inset the calculated $E(M)$
               using the fixed-spin moment method (symbols), for deformations
               of 0.0, 4.0, and 7.0\%, corresponding to SF, WF, and PM states,
               respectively.}
     \label{figure:fig11}
\end{figure*}

\section{Conclusions}

For P-graphene and N-silicene the origin of the magnetism is result of a 
partially filled very narrow band at Fermi level induced by the impurity.
This narrow impurity band causes an electronic instability which favours 
a magnetic state, which is characterized by magnetic moment of 
1.0 $\mathrm{\mu_B/cell}$ corresponding to a full-polarized impurity band. 
We found that a Stoner-type model describing the electronic instability 
of the narrow impurity band accounts for the origin of $sp$-magnetism in 
P-graphene and N-silicene. The evolution of the spin magnetic moment as a 
function of the biaxial strain is mainly governed by the behavior of the 
impurity bandwidth with the applied strain. Thus, the biaxial strain evolves 
the systems from a strong ferromagnetic state to a paramagnetic state, with 
a transition region corresponding to a weak ferromagnetic state characterized 
by a partially polarized impurity band. Furthermore, it has been demonstrated 
that with strain it is possible to modulate the spin magnetic moment and 
induce a magnetic quantum phase transition. Consequently, the control and 
manipulation of the magnetic properties in these two-dimensional magnetic 
systems are technologically attractive for the potential applications in 
spintronics and spin-based quantum computation systems, using strain 
engineering as an effective way to modulate the magnetic properties in 
2D-materials.

\section{Acknowledgments}
The authors thankfully acknowledge the computer resources, technical expertise 
and support provided by the Laboratorio Nacional de Superc\'omputo del Sureste 
de M\'exico. J.H.T. acknowledges a student fellowship from the Consejo Nacional 
de Ciencia y Tecnolog{\'\i}a (Conacyt, M{\'e}xico) and VIEP-BUAP. M.E.C.Q.
gratefully acknowledges a posdoctoral fellowship from Conacyt-M{\'e}xico.
This research was supported by Conacyt-M\'exico under grant No. 288344.

\appendix
\section{Model of narrow impurity band ferromagnetism}

According to the paramagnetic band structures at zero deformation, a narrow 
impurity band at Fermi level is present in P-graphene and N-silicene. Thus, 
the origin of magnetism could be attributed to this band because the exchange 
interaction induces a spin-splitting. Edwards and Katsnelson \cite{edwards} 
restated the conventional Stoner criteria $I N(E_F)>1$ to the case of a narrow 
impurity band where the density of states at Fermi level is approximated by 
$N(E_F) = n_{imp}/W_{imp}$, being $n_{imp}$ the impurity density is and 
$W_{imp}$ the width of the impurity rectangular band. On the other hand, 
Gruber et al. \cite{mohn-model} using a mean-field version of the Hubbard 
like interaction describe the band energy of the electronic states 
for the impurity band for each spin channel, as follows: 

\begin{equation}
E= E_{band}^{\uparrow} + E_{band}^{\downarrow} + Un^{\uparrow}n^{\downarrow}\, ,
\label{eq1}
\end{equation}

\noindent where the terms $E_{band}$ within the tight-binding approximation are

\begin{equation}
E^{\uparrow}_{band} = \int^{E_F}_{E^{\uparrow}_{min}} (E-E^{\uparrow}_c) 
                      N^{\uparrow}(E) dE \, ,
\label{eq2}
\end{equation} 

\begin{equation}
E^{\downarrow}_{band} = \int^{E_F}_{E^{\downarrow}_{min}} (E-E^{\downarrow}_c) 
                        N^{\downarrow}(E) dE \, ,
\label{eq3}
\end{equation} 
  
\noindent here $E^{\uparrow}_{c}$ and  $E^{\downarrow}_{c}$  are the center of 
the band for each spin channel which suffers an energy shift in the spin-polarized 
case with respect to the paramagnetic case, in order to have a non-zero magnetic 
moment. The $E_F$, $E^{\uparrow}_{min}$  and $E^{\downarrow}_{min}$  are the 
Fermi level and the lower limits of each band. The third term in (\ref{eq1}),  
corresponds to a Coulomb type interaction $U$ suppressing double occupancy and 
$n^\uparrow$ ($n^\downarrow$) is the number of the spin-up (spin-down) electrons 
in the band. The number of electrons in each spin band can be obtained from 
  
\begin{equation}
n^{\uparrow} = \int^{E_F}_{E^{\uparrow}_{min}} N^{\uparrow}(E) dE \, ,
\label{eq4}
\end{equation}

\begin{equation}
n^{\downarrow} = \int^{E_F}_{E^{\downarrow}_{min}} N^{\downarrow}(E) dE \, .
\label{eq5}
\end{equation}

The total number of electrons $Z$ is given by 

\begin{equation}
Z= n^\uparrow + n^\downarrow \, ,
\label{eq6}
\end{equation}

\noindent and the magnetic moment $M$

\begin{equation}
M= n^\uparrow - n^\downarrow.
\label{eq7}
\end{equation}

\begin{figure*}
\centering
\includegraphics[scale=0.6]{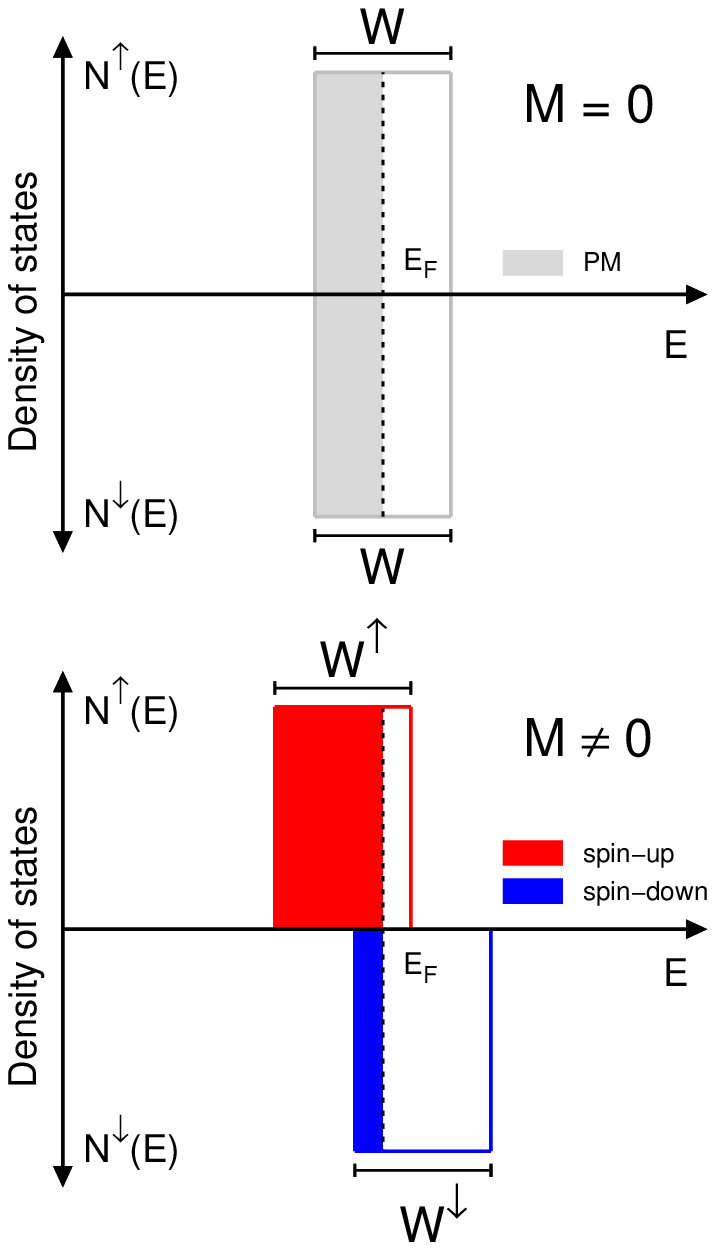}
      \caption{Density of states of the impurity band for the paramagnetic (top) 
               and spin polarized (bottom) states in the rectangular model.}
\label{figure:fig12}
\end{figure*}

Now, according to the Friedel's model \cite{friedel}, we assume a DOS with 
rectangular shape for each spin channel. In Fig.~\ref{figure:fig12} we show 
the paramagnetic and spin-polarized case according to this model. Thus, the 
density of states $N(E)$ for each spin channel is constant and is given by 

\begin{equation}
N^{\uparrow}(E) = \frac{N_0}{W^{\uparrow}} \, ,
\label{eq8}
\end{equation}
 
\begin{equation}
N^{\downarrow}(E) = \frac{N_0}{W^{\downarrow}} \, ,
\label{eq9}
\end{equation}

\noindent where $N_0$ is maximum occupancy of the band and $W^\uparrow$ 
($W^\downarrow$) is the bandwidth for each spin-up (spin-down). 

With (\ref{eq2}) and (\ref{eq8}) we rewrite the band energy for the spin-up band 

\begin{equation}
E^{\uparrow}_{band} = \int^{E_F}_{E^{\uparrow}_{min}} (E-E^{\uparrow}_c)  
                      \frac{N_0}{W^{\uparrow}} dE \, ,
\label{eq10}
\end{equation} 

\begin{equation}
E^{\uparrow}_{band} =\frac{N_0}{W^{\uparrow}}\left[\frac{(E-E^{\uparrow}_c)^2}{2}
                     \right]^{E_F}_{E^{\uparrow}_{min}} \, .
\label{eq11}
\end{equation} 

The lower limit corresponds to $E^{\uparrow}_{min} = E^\uparrow_c-(W^\uparrow/2)$ 
and the Fermi energy is obtained from (\ref{eq4}) using (\ref{eq8})

\begin{equation}
n^{\uparrow} =\left[\frac{N_0}{W^\uparrow} E\right]^{E_F}_{E^{\uparrow}_{min}} = 
              \frac{N_0}{W^\uparrow} \left[ E_F - \left( E^\uparrow_c- 
              \frac{W^\uparrow}{2} \right) \right] \, ,
\label{eq12}
\end{equation} 

\begin{equation}
E_F = \frac{n^\uparrow W^\uparrow}{N_0} + E^\uparrow_c - \frac{W^\uparrow}{2}.
\label{eq13}
\end{equation} 

With $E^{\uparrow}_{min} = E^\uparrow_c-(W^\uparrow/2)$ and $E_F$ given by 
(\ref{eq13}), we evaluate $E_{band}$ in (\ref{eq11})

\begin{equation}
 E^{\uparrow}_{band} = \frac{N_0}{2 W^{\uparrow}}
                       \left[ \left(  \frac{n^\uparrow W^\uparrow}{N_0} 
                       + E^\uparrow_c - \frac{W^\uparrow}{2}  
                       - E^\uparrow_c \right)^2 - \left(  E^{\uparrow}_c 
                       - \frac{W^\uparrow}{2} -E^\uparrow_c \right)^2 \right]
\label{eq14}
\end{equation} 

\noindent and simplifying, we arrive to:

\begin{equation}
E^{\uparrow}_{band} =\frac{W^{\uparrow}}{2} \left[ \frac{n^{{\uparrow}2}}{N_0} 
                     - n^\uparrow  \right].
\label{eq15}
\end{equation} 

In the same way, it is found that $E^\downarrow_{band}$ is: 

\begin{equation}
E^{\downarrow}_{band} =\frac{W^{\downarrow}}{2} \left[ \frac{n^{{\downarrow}2}}
                       {N_0} - n^\downarrow  \right].
\label{eq16}
\end{equation} 

Replacing the expressions given in (\ref{eq15}), (\ref{eq16}) of  
$E^{\uparrow}_{band}$ and $E^{\downarrow}_{band}$ in (\ref{eq1}), 
we obtain that  

\begin{equation}
E = \frac{W^\uparrow}{2} \left( \frac{n^{\uparrow2}}{N_0} - n^\uparrow \right) 
  + \frac{W^\downarrow}{2} \left( \frac{n^{\downarrow2}}{N_0} - n^\downarrow \right) 
  + Un^{\uparrow}n^{\downarrow}. 
\label{eq17}
\end{equation}
  
Now, from  (\ref{eq6}) and (\ref{eq7}), $n^{\uparrow}$ and $ n^{\downarrow}$ can be 
expressed in terms of the magnetic moment $M$ and the total number of electrons $Z$

\begin{equation}
n^{\uparrow} = \frac{Z+M}{2} \, ,
\label{eq18}
\end{equation}

\begin{equation}
n^{\downarrow} = \frac{Z-M}{2}. 
\label{eq19}
\end{equation}

Using this expressions for $n^{\uparrow}$ and $n^{\downarrow}$ in (\ref{eq17}), 
we obtain an expression for the energy $E$ as a function of $M$

\begin{eqnarray}
E(M) & = & \frac{Z^2}{4} \left( \frac{W^\uparrow+ W^\downarrow}{2N_0} + U \right) 
          -\frac{Z}{4}(W^\uparrow + W^\downarrow) \nonumber \\
     & + & \frac{M}{4}(W^\uparrow - W^\downarrow) \left( \frac{Z}{N_0} -1 \right)  \\
     & + & \frac{M^2}{4}\left( \frac{W^\uparrow+ W^\downarrow}{2N_0} - U  \right) \nonumber .
\label{eq20}
\end{eqnarray}

The first two terms are independent of $M$, the third term is a linear dependence 
of $M$, however in the case of half-filling ($N_0 = Z$) this term vanishes. 
The last term has a quadratic dependence of $M$ and for spontaneous magnetization 
needs to be less than 0:

\begin{equation}
 \frac{W^\uparrow+ W^\downarrow}{2N_0} - U  < 0, 
\label{eq21}
\end{equation}

\noindent which can be written as  

\begin{equation}
	\frac{2N_0U}{W^\uparrow+ W^\downarrow} > 1.
\label{eq22}
\end{equation}

For the case of a rigid band splitting ($W^\uparrow = W^\downarrow$), 
the condition for  spontaneous magnetization is: 

\begin{equation}
	U\frac{N_0}{W} > 1. 
\label{eq23}
\end{equation}

We can see that this condition is a Stoner-like criterion, since the ratio 
$N_0/W$ is the density of states per spin channel in the present model. 
Thus, it is clear that for a narrow impurity band where $W$ is very small, 
this condition can be fulfilled even for small or moderate $U$ values and 
low occupancy.  The present model should be suitable for systems that present a full-polarized impurity band, i.e. in the strong ferromagnetims regime.

\end{document}